\documentclass[conference,10pt]{IEEEtran}

\usepackage{amsmath}
\usepackage{amssymb}
\usepackage{booktabs}
\usepackage{url}
\usepackage{xspace}
\usepackage{balance}
\usepackage{hyperref}

\usepackage{graphicx}
\usepackage{subcaption}
\usepackage{circledsteps}
\usepackage{multirow}
\usepackage{array}

\usepackage{tcolorbox}

\usepackage{listings}
\usepackage{xcolor}

\lstdefinelanguage{PTXcuda}{
  morekeywords={SYNCS, PHASECHK, TRANS64, TRYWAIT, NANOSLEEP, BRA},
  morecomment=[l]{//},
}

\usepackage{cleveref}
\usepackage{svg}

\begin{document}

\title{Architecting the Next Generation of Asynchronous, Distributed GPUs for the AI Era}

\def\hpcacameraready{}

\newcommand{\hpcaauthors}{%
  Junrui Pan,
  Weili An,
  Cesar Avalos Baddouh,
  Ni Kang,
  Christin David Bose,
  Aaron Barnes,\\
  Ahmad Alawneh,
  Fangjia Shen,
  Yechen Liu,
  Anusuya Nallathambi,
  Atthin Chandrashekar,
  Timothy G. Rogers%
}
\newcommand{\hpcaaffiliation}{Purdue University, West Lafayette, IN, USA}
\newcommand{\hpcaemail}{%
  \{pan251, an107, cavalosb, kang222, chris241, barnes88,
  aalawneh, liu2550, anallat, chand158, timrogers\}@purdue.edu,\\
  fangjias@acm.org%
}

\ifdefined\hpcacameraready
  \author{%
    \IEEEauthorblockN{\hpcaauthors}
    \IEEEauthorblockA{\hpcaaffiliation\\\hpcaemail}
  }
\else
  \author{\IEEEauthorblockN{Paper \#1596}}
\fi


\maketitle

\begin{abstract} \label{sec:abstract}
The rapid evolution of machine learning workloads has fundamentally transformed GPU hardware, driving architectures toward Multi-Chip Module (MCM) topologies, asynchronous execution primitives, and persistent, multi-phase kernel behaviors. Despite these shifts, cycle-level simulation infrastructure has lagged behind, lacking the native capability to model the physical non-uniformity of modern GPUs alongside the massive scale of state-of-the-art AI workloads. To bridge this gap, we present a cycle-level simulation framework designed to accurately model modern GPU generations, including Ampere, Hopper, and Blackwell. Rigorously validated against physical silicon, the simulator achieves a 99\% Pearson correlation coefficient and a 13.5\% mean absolute cycle error on the H100 GPU. Utilizing this infrastructure, we conduct architectural case studies to evaluate emerging design trajectories, including chiplet topology scaling, expanded SRAM capacity and bandwidth, and inter-GPU prefetching strategies.
\end{abstract}

\section{Introduction} \label{sec:intro}
Machine learning has altered the trajectory of Graphics Processing Unit (GPU) design. While datacenter GPUs continue to support traditional High Performance Computing (HPC) workloads, the sheer scale of modern AI has forced a paradigm shift. Contemporary GPUs are now fundamentally designed to accelerate machine learning training and inference. However, unlike the myriad of machine-learning specific acceleration hardware, GPUs remain committed to maintaining a level of programmability that allows them to adapt to changes in the algorithms. As a result, the design of GPUs is an interesting case study in the tensions between general programmability and efficiency. The last decade of GPU hardware and software design provides many examples, from the introduction of bulk CISC-like Tensor Memory Accelerators (TMAs) and Tensor Cores, to the introduction micro- and multi-GPU for scaling, to the software optimizations in kernel fusion, warp specialization, asynchronous in-core programming, persistent kernels and more. The way GPUs are programmed for peak performance has changed dramatically in the last decade. This shift has opened up new questions facing both contemporary and future GPU designs. In this paper, we identify a set of meaningful architectural questions on the future and current state of GPUs and introduce Accel-Sim 2.0, a validated, credible simulation framework to answer them.


The demand for AI compute capacity has created intense competition in the hardware accelerator space. Startups and academic projects have proposed diverse, highly-specialized designs that trade the programmability of the GPU for increased efficiency \cite{rocki_fast_stencilcode_2020,abts_softwaredefined_tensor_2022,google_tpu_2017,amazon_trainium_2025,TensorTorrent_2025,meta_MITA_2023,scnn_2017,cambricon-D_2025,FACT_2023}. For example, the emergence of inference as a dominant workload has highlighted the advantages of SRAM-heavy chips such as Cerebras' Wafer-Scale Engine \cite{rocki_fast_stencilcode_2020}, which provisions 44 GB of SRAM to deliver extreme on-chip bandwidth, and the Groq (now NVIDIA) LPU \cite{abts_softwaredefined_tensor_2022}, another SRAM-heavy chip that relies on software-defined determinism to eliminate hardware synchronization overhead. These designs represent significant architectural departures from traditional GPUs. Despite the clear advantage in bandwidth provided by SRAM and the intuitive advantage of static scheduling (in the case of the LPU), it is unclear how much the GPU's complex microarchitecture and dynamic, non-deterministic timing are hindering its performance. If a GPU with much more bandwidth can be built without changing the software, ISA, synchronization mechanisms, or core configuration, would that be enough? How much is asynchronous costing GPUs \cite{jooybar_gpudet_}? 
By accurately modeling the asynchronous primitives and hardware-orchestrated data movement of contemporary architectures, we leverage our newly developed framework to push GPU off-core bandwidth to hypothetical limits, isolating shifting execution bottlenecks to reveal that the synchronization overheads typically targeted by deterministic architectures are effectively masked by contemporary GPU designs (Section~\ref{sec:gpu-v-lpu}). 

\begin{figure*}
    \centering
    \includegraphics[width=\linewidth]{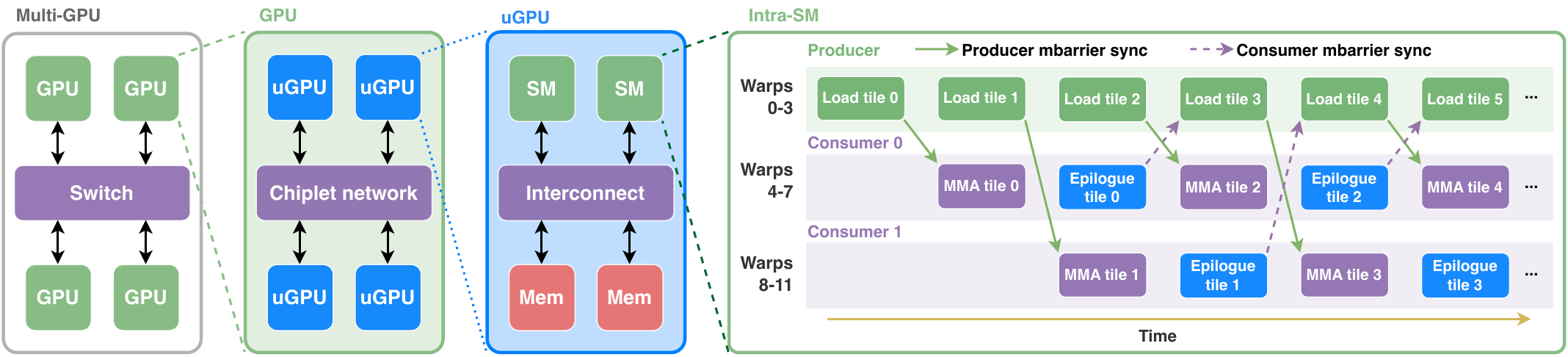}
    \caption{GPU execution hierarchy from multi-GPU to multi-die chiplet to intra-SM warp specialized kernel pipeline with producer-consumer synchronization via mbarrier.}
    \vspace{-10pt}
    \label{fig:warpspecialized_gemm}
\end{figure*}

Another major trend in GPU design is driven by the end of Moore's Law and silicon reticle limits~\cite{arunkumar_mcmgpu_multichipmodule_2017,shalf_future_computing_2020, hennessy_new_golden_2019}. To scale compute, GPU systems are disaggregated within the package into micro-GPUs on different chiplets. 
While the programming model maintains the illusion of a single flat device to ease developer burden, this introduces cache coherence and Non-Uniform Memory Access (NUMA) effects.
By contrasting our faithful model of a contemporary GPU with a hypothetical GPU unhindered by excessive crossbar scaling and fewer transistors, we ask if this design decision matters. Do transparent micro-GPUs hurt performance versus a monolithic design, or are the effects hidden? Our evaluation reveals that LLMs are naturally resilient, as prefill phases are compute-dense enough to absorb the added inter-chiplet latency, and the decode phases lack the locality to be penalized by the loss in L2 capacity. However, capacity-sensitive workloads relying on spatial reuse in the L2 suffer. Due to data replication shrinking the effective cache capacity, graph analytics workloads such as MST degrade by 23\% and ResNet-50 by 20.7\% compared to a hypothetical monolithic design. Furthermore, even LLMs pay a hidden tax, evaluating them against a monolithic design introduces 17\% more L2 misses, increasing energy consumption in the memory system (Section~\ref{sec:micro-gpu}).

Finally, as AI models eclipse single-device memory capacity limits, evaluating multi-GPU scale-out architectures becomes essential. 
This problem is especially pronounced in capacity-constrained Deep Learning Recommendation Models (DLRMs) that require lookups from large tables no single GPU can hold.
To investigate these inter-GPU dynamics, we construct a case study exploring the use of Unified Virtual Addressing (UVA) to expand the effective capacity of DLRMs from Meta~\cite{metasyn}. To overcome the latency of implicit remote memory accesses over NVLink, we evaluate a UVA-based remote prefetching mechanism. By accurately modeling the underlying interconnect topologies, our evaluation reveals that while the proposed prefetching strategy accelerates distributed lookups, overall performance plateaus. 
More importantly, we demonstrate that to fully exploit inter-GPU bandwidth, future architectures must decouple memory prefetch depth from the strict capacity limits of in-core resources (Section~\ref{sec:multi-gpu}).

Uncovering these architectural insights requires visibility into the behavior of contemporary systems and a simulation model that credibly supports the diverse architectures and range of workloads that run on GPUs.
No existing open-source simulator is capable of executing the complex, asynchronous kernels that comprise contemporary AI frameworks like PyTorch and vLLM \cite{kwon_efficient_memory_2023}, or optimized implementations like Flash Attention~\cite{dao_flashattention_fast_2022,dao_flashattention2_faster_2023,shah_flashattention3_fast_2024}, NVIDIA's cuBLAS~\cite{cublas} or CUTLASS~\cite{thakkar_cutlass_2023}.

The most closely related infrastructures that target NVIDIA GPUs are GPGPU-Sim \cite{bakhoda_analyzing_2009} and Accel-Sim 1.x\cite{khairy_accelsim_2020}, yet both lack contemporary hardware modeling and multi-GPU support. 
To overcome these visibility limitations and enable our cycle-level case studies, we extensively update the Accel-Sim 1.x framework to support the validated simulation of any kernel that runs on NVIDIA architectures from Volta through Blackwell. 
In addition to cycle-level simulator updates, we discovered that understanding the behavior of complex fused, persistent kernels requires time-series, fine-grained in-silicon profiling that standard profiling tools like Nsight Systems and Nsight Compute cannot easily provide. To gain this kind of visibility into profiled GPU results, we developed GPUVision, a CUPTI-based \cite{nvidia_cuda} profiling tool that collects microarchitectural time-series metrics from real hardware, enabling high-fidelity comparison against simulation outputs. We validate Accel-Sim 2.0 against NVIDIA H100 and B200 hardware using over 34,000 kernel instances each across 22 benchmark suites, achieving a 0.99 Pearson correlation coefficient and 13.5\% mean absolute error on cycles on Hopper and 8.9\% on B200. (detailed evaluation in Section~\ref{evaluation})

This paper makes the following contributions:

\begin{itemize}
\item Accel-Sim 2.0, a comprehensive cycle-level GPU simulation framework for Hopper and Blackwell architectures, paired with a novel CUPTI-based \cite{nvidia_cuda} profiling tool, GPUVision. Together, they enable high-fidelity execution and hardware validation of state-of-the-art workloads, natively capturing both emerging hardware mechanisms (e.g., multi-chip topologies, threadblock clusters) and the dynamic temporal phasing obscured by traditional aggregate metrics.
\item A comprehensive set of case studies evaluating future GPU design trajectories, specifically analyzing the performance implications of expanding memory architectures and the critical microarchitectural impact of asynchronous synchronization overheads.
\item A detailed analysis of NUMA effects on GPUs, demonstrating that while inherent latency tolerance protects LLM workloads, diminished effective cache capacity hurts performance in locality-dependent applications like graph analytics and CNNs.
\item A case study on optimizing bottleneck kernels in deep learning recommendation models using our multi-GPU simulator. In particular, we evaluate the implications of prefetching memory accesses over the inter-GPU interconnect enabled by Unified Virtual Addressing (UVA).
\end{itemize}

\section{{Accel-Sim 2.0}}
{
The NVIDIA Hopper marks a fundamental shift in how GPUs are programmed, and no existing public microarchitecture simulator (including Accel-Sim 1.x) is capable of modeling the core features (Table \ref{tab:sim_config}). To quantify this limitation, we ran Accel-Sim 1.x scaled to resemble an H100. We mapped unsupported opcodes to the nearest equivalent execution unit (e.g., \textit{WGMMA} to the HMMA unit) or replaced them with NOPs when no equivalent instruction exists (e.g., \textit{TMA}). The results demonstrate severe inaccuracies across two distinct workload domains. First, for SIMT-heavy HPC workloads (Table \ref{tab:hpc-workload}), Accel-Sim 1.x yields a 36\% mean absolute percentage error in GPC cycles, alongside massive deviations in the memory hierarchy (85\% error for L2 read accesses and 59\% for L1 reads). Second, when evaluating MMA-dominant, warp-specialized CUTLASS GEMMs \cite{thakkar_cutlass_2023}, the baseline yields a prohibitive 46\% cycle error. Crucially, because 1.x entirely lacks support for the TMAs, the resulting memory subsystem metrics are architecturally invalid. Furthermore, the lack of support for \textit{mbarrier} synchronizations creates completely unrealistic execution patterns that ignore dependencies. Because the legacy simulator cannot expose the true bottlenecks of contemporary systems, our work is not an incremental accuracy improvement; rather, it enables architectural studies that were previously impossible. 
}
\subsection{{Contemporary GPU Programming Models}} \label{comteporary_gpu}
{Traditionally, GPGPU computing relied on a bulk synchronous SIMT paradigm in which all warps within a threadblock issue shared memory loads concurrently and then synchronize at a block-level barrier before computation proceeds. Meeting the throughput demands of contemporary AI workloads, however, requires departing from this independent-thread execution model in favor of orchestrating tightly coupled, asynchronous tasks.
}

{
Warp specialization~\cite{crago_wasp_2024,bauer_cudadma_optimizing_2011} has become the standard approach for high-performance GPU kernels (Figure~\ref{fig:warpspecialized_gemm}, right). It decouples execution by dedicating producer warps to fetching data from global memory while consumer warps concurrently perform matrix math and epilogue processing on previously loaded tiles. 
On contemporary Hopper and Blackwell architectures, such cycle-level orchestration is accelerated by dedicated hardware units, including the TMA, Asynchronous Warp Group Level operations (\texttt{WGMMAs}) for Hopper and \texttt{UTCMMA} for Blackwell. Synchronization between these operations is performed through \texttt{mbarrier} primitives.
This asynchronous model is now pervasive, underpinning FlashAttention (FA)~\cite{shah_flashattention3_fast_2024,zadouri_flashattention4_algorithm_2026} as well as core NVIDIA libraries such as NCCL~\cite{NCCL}, cuSolver~\cite{cusolver2026}, and CUTLASS~\cite{thakkar_cutlass_2023}.
}

{
TMA enables hardware-orchestrated bulk data movement. Because the TMA uses the uniform programming model, a single thread can issue a TMA instruction to initiate the transfer of a multidimensional data tile. The TMA hardware then independently generates the memory requests and manages the transfer, decoupling data movement from the compute pipeline and reducing instruction issue overhead. Alongside the TMA, Hopper introduces threadblock clusters (Cooperative Groups, CGAs), which let multiple SMs share a distributed shared memory pool. Building on this, TMA multicast allows a single thread to broadcast one memory tile to all SMs in a cluster, further increasing effective bandwidth.
Asynchronous execution between units requires fine-grained synchronization via the \texttt{mbarrier} (\texttt{SYNCS}), a 64-bit shared-memory object that tracks both thread arrivals and outstanding asynchronous memory transfers. This transaction count can be updated by either a TMA or a warp-issued \texttt{LDGSTS} operation. Consumer warps synchronize with \texttt{mbarrier.try\_wait}, repeatedly polling the barrier's phase bit in a software spinloop that often yields via \texttt{nanosleep}.
The trace below shows a representative polling loop from FA-3 on Hopper. Although drawn from FA-3, this pattern is essentially the same across every \texttt{mbarrier}-based kernel we examined.
}
\vspace{-3pt}
\begin{tcolorbox}[
    colback=white!15,     
    colframe=black,      
    arc=3mm,             
    boxrule=0.5pt,       
    left=6pt, right=6pt, top=0pt, bottom=0pt, 
    width=\linewidth,    
]
{\footnotesize c1f0 SYNCS.PHASECHK.TRANS64.TRYWAIT // mbarrier}

{\footnotesize c200 NANOSLEEP // nanosleep}

{\footnotesize c210 SYNCS.PHASECHK.TRANS64 // predicate for BRA}

{\footnotesize c220 BRA c1f0 // jump to start of spinloop}
\end{tcolorbox}
{The \texttt{mbarrier}s are confined to cluster scope. The same mechanism can instead be implemented with global loads, closely resembling the \texttt{mbarrier} loop above except that the control data is fetched via \textit{LDG.E.STRONG.SYS}; the \texttt{STRONG.SYS} qualifier indicates that the data is served from L2, bypassing L1.
This software-driven polling severely compromises trace-driven simulation. The exact number of spinloop iterations executed depends on hardware timing, which is inevitably inflated by the overhead of the tracing process. As a result, the trace captures an excessive and unrepresentative number of polling iterations. Treating these dynamic polling loops as static architectural artifacts is incorrect, and replaying them distorts the simulated execution timing.
}
The Tensor Core on Hopper can leverage warpgroups to perform matrix multiplications on larger tiles asynchronously. A warpgroup consists of four contiguous warps, each scheduled onto one subcore, and the four subcores cooperate on a single WGMMA operation. 
Before the operation begins, a barrier (\textit{WARPGROUP.ARRIVE}) synchronizes the warps to ensure every warp is ready, and a second barrier (\textit{WARPGROUP.DEPBAR}) confirms that all warps have completed. Neither this fine-grained synchronization nor the warpgroup abstraction was previously supported. Both are implemented in Accel-Sim 2.0 to accurately capture Hopper Tensor Core throughput.

\subsection{{Tracer}} \label{tracer}
We build our tracer on TMA-enabled NVBit v1.8 \cite{NVBIT}. To support Tensor Descriptors, which are used by \textit{mbarrier}s and \textit{TMA}s, we implement dynamic register value tracing. Although our implementation supports generic register tracing across all instructions, we selectively enable it only for critical synchronization primitives to minimize tracing overhead. We adopt a spinloop detection methodology inspired by NVAS~\cite{villa_need_2021}: during trace generation, we run the kernel twice and compare instruction counts between passes to isolate non-deterministic spinloop sections. The tracer then filters out redundant iterations, emitting a single canonical pass of the loop.
During simulation, rather than blindly replaying static traces, we functionally model spinloops and evaluate synchronization dynamically. Upon a failed asynchronous wait, the simulator natively reproduces the hardware's structural behavior (the spinloop). This dynamic evaluation faithfully reproduces true hardware synchronization behavior, enabling the detailed overhead studies presented in Section~\ref{sync_overhead}.

\begin{table*}[h]
\caption{{Hopper H100 configuration, contrasting the Accel-Sim 1.x baseline with features added in Accel-Sim 2.0.}}
\label{tab:sim_config}
\resizebox{\textwidth}{!}{%
\begin{tabular}{|l|c|c|}
\hline
                           & \textbf{Inherited from Accel-Sim 1.x}    & \textbf{Accel-Sim 2.0}                                                                                                                                                                                  \\ \hline
\# SMs                     & 132                                      & \textbf{\begin{tabular}[c]{@{}c@{}}new operand collector, spinloop support, chiplet-aware CTA scheduling, \\ Cooperative Groups support, cluster control\end{tabular}}                                  \\ \hline
L1 Cache/Shared Memory     & 256\,KB, 4 banks                         & \textbf{TMA, CGA multicast, Distributed Shared Memory}                                                                                                                                                                             \\ \hline
\# Exec Units$^\dagger$    & 4 FPs, 4 DPs, 4 INTs, 4 SFUs, 4 HMMAs    & \textbf{WGMMA, UTC*MMA, warpgroup commit/wait, 2-CTA, exec unit refactor}                                                                                                                               \\ \hline
L2 Cache                   & 50\,MB, 80 Banks                         & \textbf{modulo IPOLY hash, LRC, chiplet cache policies}                                                                                                                                                 \\ \hline
MMA Latency                & Fixed                                    & \textbf{Variable (depending on N)}                                                                                                                                                                      \\ \hline
Memory                     & HBM bandwidth/latency                    & \textbf{HBM3, HBM3e bandwidth/latency}                                                                                                                                                                  \\ \hline
Interconnect               & Monolithic crossbar                      & \textbf{chiplet interconnect, NVLink}                                                                                                                                                                   \\ \hline
Synchronization Primitives & Barrier                                  & \textbf{mbarrier, bar.arv, bar.sync, ldgstsbar, fence, UTCBAR, WARPGROUP}                                                                                                                                          \\ \hline
Tracer                     & SASS register dependencies, control flow & \textbf{\begin{tabular}[c]{@{}c@{}}register value tracing, spinloop handling, multi-GPU tracing, trace compression, \\ trace paging, tensor descriptor support, Pytorch per-layer support\end{tabular}} \\ \hline
Performance Counters       & 54                                       & \textbf{11,072}                                                                                                                                                       \\ \hline
Correlation                & kernel-level Nsight Compute              & \textbf{cycle-level GPUVision}                                                                                                                                                                          \\ \hline
\end{tabular}%
}
\footnotesize{$^\dagger$FP: single-precision floating point, DP: double-precision floating point, INT: integer, SFU: special function unit, HMMA: Volta tensor core.}
\end{table*}

To support multi-GPU workloads, we extend the tracer into an NVLink-aware, multi-context tracer that captures all architectural events for a given GPU context along with the CUDA memory-management metadata needed to reason about allocation ownership and sharing. 
Because addresses are assigned at runtime, the tracer uses NVIDIA driver APIs (e.g., \textit{CudaMemMap}) to build a precise per-GPU virtual address map, which it passes to the simulation engine.
During simulation, the framework consults this map to classify each memory request as a local or a remote access, routing remote requests through the interconnect model via RDMA. The interconnect is a custom subsystem based on BookSim~\cite{booksim}; it natively supports common topologies such as rings, switch-based clusters, and all-to-all networks, while its modular configuration files can be easily adapted to emerging or proprietary topologies. The tracer additionally records the inter-kernel synchronization events between CPU-GPU and/or GPU-GPU, which the simulator frontend uses to construct a Directed Acyclic Graph (DAG) of kernel dependencies.

To address the severe storage bottlenecks inherent to trace-driven simulation, we implement native \textit{zstd} compressed trace I/O, achieving an average storage reduction of 27.2$\times$ across our workload suite.
Accel-Sim 1.x loads traces at CTA granularity. This becomes a serious problem under persistent kernel execution (commonly used by frameworks like vLLM \cite{kwon_efficient_memory_2023}), where the per-CTA trace grows enormously; in such cases, we observed the simulator frequently requiring up to 200 GB of runtime memory. To resolve this, we introduce page-based trace loading. For each warp within each CTA, the trace is compressed into chunks (2048 instructions by default, chosen to balance compression ratio against memory footprint), and the trace is double-buffered so that the simulator streams instructions chunk by chunk. With this scheme, runtime memory remains bounded at roughly 4 GB regardless of kernel size.
Tracing every kernel in an LLM is prohibitively expensive, both in tracing time and in the volume of traces produced. Fortunately, machine learning workloads exhibit highly predictable kernel launch sequences. This regularity is especially pronounced in LLMs, which are composed of identical, repeating transformer blocks, so tracing a single block provides sufficient data to accurately extrapolate end-to-end performance~\cite{avalosbaddouh_principal_2021}. To exploit this, we implement a PyTorch hook that dynamically toggles the NVBit tracer, letting us flexibly select exactly which layers to trace.
To validate that a single transformer block accurately extrapolates to the full model on real hardware, we profiled a complete Llama-3.1-8B inference pass (2K-token prefill, 128 decode steps) on an H100 using Nsight Systems. By placing NVTX ranges across all 32 decoder blocks, we captured every primary and auxiliary kernel without omission; consequently, multiplying the measured latency of a single block by 32 reproduces the directly measured full-stack hardware latency within 0.97\% for prefill and 3.06\% for decode.

\subsection{{Memory Subsystem}} \label{memory_subsystem}
While the Blackwell architecture formalizes multi-die architecture, preceding architectures (Ampere and Hopper) already exhibit spatial disaggregation through logical L2 partitioning. A uGPU (Figure \ref{fig:warpspecialized_gemm}) is a logical partition of a single GPU die that behaves like a chiplet: each uGPU owns private L2 partitions and directly attached High Bandwidth Memory (HBM) stacks. This structural division introduces pronounced NUMA effects that fundamentally alter contemporary GPU performance profiles \cite{choudhary_optimizing_attention_2025}.

\begin{figure}[h]
    \centering
    \includegraphics[width=\linewidth]{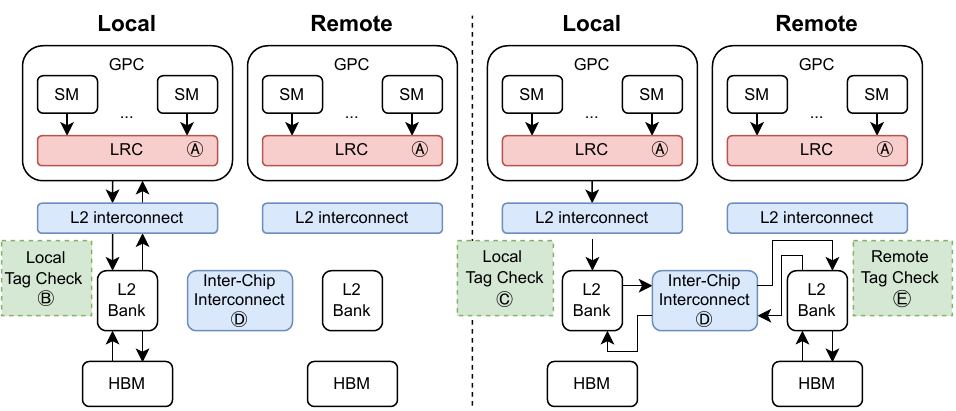}
    \caption{High-level overview of the modeled micro-GPU memory subsystem, derived from targeted microbenchmarking of the NVIDIA H100 and B200 architecture.}
    \label{fig:mem_subsystem}
\end{figure}

Figure \ref{fig:mem_subsystem} illustrates the partitioned L2 cache architecture modeled in this work to capture these NUMA dynamics. To reduce latency, Hopper and subsequent architectures replicate data across partitions to mitigate interconnect overhead \cite{jin_uncovering_real_2024}.
To reduce the traffic entering this distributed L2, Hopper employs an L2 Request Coalescer (LRC) \textcircled{\scriptsize{A}} to aggregate L1 misses from different SMs targeting the same L2 sector. Consequently, L2 cache traffic is not proportional to L1 miss volume. Through microbenchmarking, we characterize the LRC as a per-GPC structure.
We evaluate two configurations: two CTAs executing on the same GPC, and two CTAs executing on different GPCs, using Cooperative Groups to enforce the intended placement. We find that the LRC coalesces up to four read requests into a single transaction (a 4:1 merge ratio) exclusively when those requests originate from SMs within the same GPC. 
Following this coalescing stage, the GPC routes all memory requests to the L2 banks residing within the same spatial partition (local L2). 

While NVIDIA's hardware implementations remain publicly undisclosed, we conducted targeted microbenchmarking to study these local and remote memory behaviors. 
We begin by characterizing how data is mapped across uGPUs using a memory latency test. The microbenchmark confines execution to a single warp and issues 128B strided loads. The resulting latency exhibits a clear periodic pattern: each 8192B region alternates between slow and fast accesses. This reveals a uGPU mapping stride of 8192B, matching that of the A100~\cite{jia2021dissecting}. Repeating the experiment across every SM identifies which SMs are co-located on the same uGPU. We then select two SMs residing on distinct uGPUs to probe the L2 cache policy. 
Our L2 cache policy microbenchmarks results show that the memory subsystem operates as shown in Figure \ref{fig:mem_subsystem}: if the target data resides in the local HBM, the local L2 fetches it directly upon a miss \textcircled{\scriptsize{B}}, bypassing the inter-chip interconnect. If the data maps to the remote HBM, the request must still perform a local tag check \textcircled{\scriptsize{C}} to determine if a replicated copy already resides in the local L2. If this local check misses, the request is forwarded via the inter-chip interconnect \textcircled{\scriptsize{D}} to the remote L2 partition. The remote L2 then performs its own tag check; a hit \textcircled{\scriptsize{E}} returns the data across the interconnect, while a remote miss triggers a fetch from the remote HBM. Once the remote access is resolved, the retrieved data is replicated and cached in both L2 partitions. 

While CUDA relies on a relaxed memory consistency model rather than strict inter-thread ordering \cite{lustig_formal_analysis_2019,muthukrishnan_gps_global_2021,NUMAgpu,ren_hmg_extending_2020}, the hardware must guarantee that a store committed to a specific L2 partition is accessible to read requests originating from any SM across the different uGPUs.
Our microbenchmarks reveal that the hardware uses a hybrid “local write-through, remote write-back” policy. We select two SMs on different uGPUs using the methodology described above. One SM writes to an address, after which both SMs load it. Both loads hit locally, and no inter-uGPU load traffic is observed in Nsight Compute.
This shows that all store requests initially route to the local L2. Regardless of the underlying physical HBM mapping, the local L2 performs a write-through by forwarding the data across the interconnect \textcircled{\scriptsize{D}} to the remote L2. The remote L2 then handles the request using a standard write-back policy. To prevent conflicting memory updates during cache line replacement, write-backs to the HBM are arbitrated by physical address ownership: only the L2 partition corresponding to the physical home node writes dirty data back to memory, while the non-owner partition discards it.

\subsection{{Simulator}}
Table \ref{tab:sim_config} highlights the changes we made on top of the baseline Accel-Sim 1.x framework. Besides enabling new critical features used in contemporary LLM workloads, such as TMA and new asynchronous primitives, we updated the operand collector for both performance and accuracy as observed by previous works \cite{huerta_dissecting_2025}. With other software optimizations such as SIMT pipeline refactoring, we achieved 27.5k kilo warp instructions per second, a 2.2x improvement over Accel-Sim 1.x, even with all the new features. 
To support a non-power-of-two number of L2 banks (H200s, B200s), we implement a combined IPOLY+MODULO hash. 
Beyond Hopper, Blackwell datacenter GPUs introduce TCgen5 tensor cores (\texttt{UTCMMA}) and a cooperative 2-CTA execution mode where adjacent thread blocks process larger tile sizes, all of which are modeled with cycle-level fidelity in our framework.

Accel-Sim 1.x ships with scripts that collect kernel-level statistics through Nsight tools and perform kernel-level correlation. 
To gain additional insights into the hardware, we develop GPUVision, which profiles and graphs hardware GPU performance metrics as a function of cycle count. We emphasize that GPUVision is \emph{not} based on simulation: it measures real hardware. It leverages CUPTI's Performance Monitor sampling to collect performance counters. GPUVision exposes the same metric set as Nsight Compute; any counter available in NCU can be collected. The difference is that GPUVision reports these metrics as a cycle-level timeseries in CSV format rather than a single aggregated value per kernel. Compared with NVIDIA Nsight Compute and Nsight Systems, GPUVision generates easily consumable CSV output, and it's able to hook arbitrary CUDA API calls. GPUVision profiles events that Nsight Compute cannot; for example, Nsight Compute is restricted to kernel launches, whereas GPUVision can also profile APIs such as \texttt{cudaMemcpy}.
The real strength comes from pairing the hardware GPUVision with the reworked simulation statistics. Researchers can perform cycle-level correlation between measured and simulated execution (Figure~\ref{fig:fa3_timeseries}), unlocking a level of analysis that was previously impossible.

\section{Deconstructing Asynchrony}
\label{sec:gpu-v-lpu}
To demonstrate the architectural studies enabled by Accel-Sim v2.0, we explore the broader landscape of AI acceleration. Within this domain, the Groq LPU \cite{abts_softwaredefined_tensor_2022} and contemporary GPUs represent two diverging design paradigms: software-defined, deterministic scheduling with massive on-chip SRAM versus complex, hardware-managed asynchronous execution. To quantify the architectural tradeoffs between these approaches and guide future trajectories, we evaluate optimized, warp-specialized CUTLASS GEMM kernels \cite{thakkar_cutlass_2023} on the validated H100 baseline configuration. This setup isolates the exact coordination costs of execution by incorporating two key characteristics of emerging custom accelerators. First, we scale up L2 bandwidth to determine the precise threshold required to eliminate memory transfer as the primary execution bottleneck. 
Second, we evaluate the synchronization overhead inherent to dynamic GPU execution by isolating the latency of asynchronous coordination primitives (such as \texttt{mbarrier}). By comparing contemporary runtime scheduling against a deterministic baseline free of barrier synchronization costs, we quantify the performance tax of dynamic coordination to determine the true impact of asynchronous primitive overhead on end-to-end throughput.

\subsection{Unconstrained Bandwidth} \label{bw_expansion}
To explore the performance bounds of such a paradigm, we ensure L2 always hits and extend available L2 bandwidth by overclocking the memory subsystem. Crucially, we maintain the baseline throughput for the SM's internal shared memory, register file, and functional unit pipelines. Because specialized MMA warps must still load operands through the internal shared memory, any observed performance improvements result from alleviating the global, off-SM bandwidth bottleneck rather than accelerating SM-internal execution.

\begin{figure}[h]
    \centering
    \includegraphics[width=\linewidth]{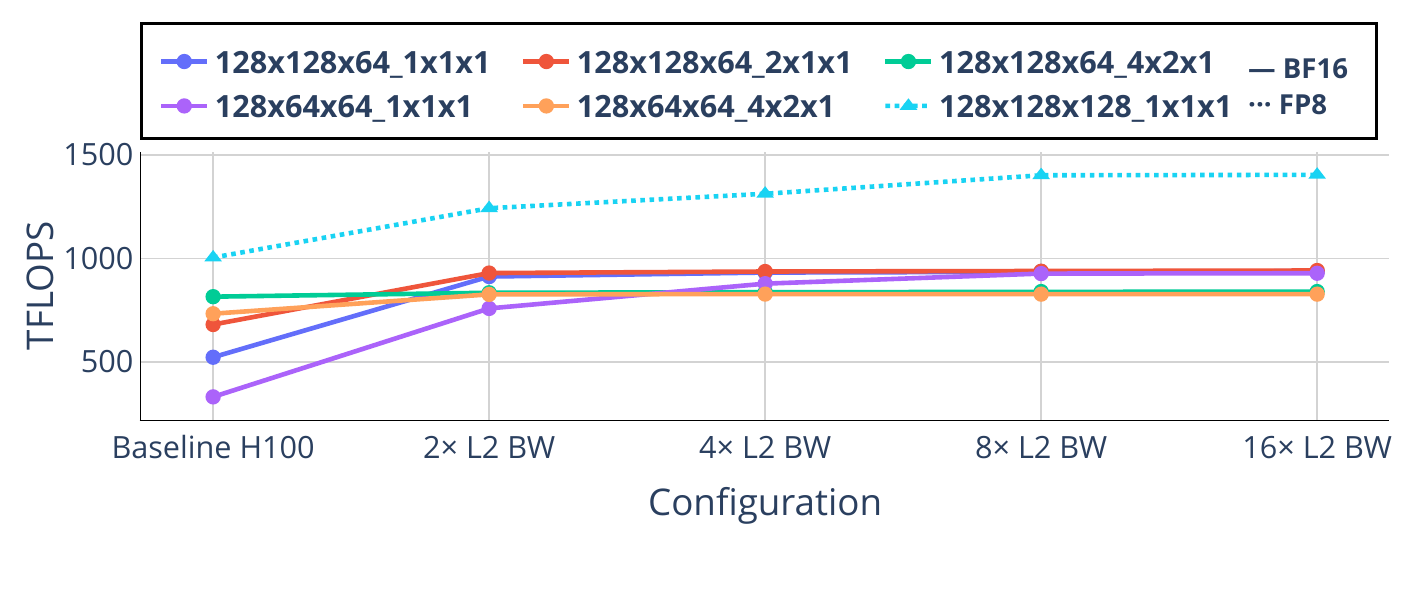}
    \caption{Simulated H100 TFLOPS for a 4096x4096x4096 dense GEMM evaluated with CUTLASS. Kernel configurations are denoted as [Threadblock Tile Shape (MxNxK)] – [Cluster Dimension].}
    \label{fig:async_bw_sweep}
\end{figure}

We evaluate a 4096x4096x4096 dense GEMM workload using the CUTLASS Profiler \cite{thakkar_cutlass_2023}, sweeping across various tile sizes and threadblock cluster dimensions. Figure \ref{fig:async_bw_sweep} illustrates the resulting performance scaling as a function of L2 bandwidth.
Three distinct scaling trends emerge from this evaluation.
First, expanding L2 bandwidth is disproportionately effective for smaller threadblock clusters.
Large clusters (e.g., 4x2x1) use TMA multicast: CTAs in the same cluster row or column consume the same input tile, so a single TMA request fetches the tile once and broadcasts it into the shared memory of every participating CTA, reducing aggregate L2 bandwidth demand.
These configurations are therefore compute-bound at baseline and gain little from the expanded L2 bandwidth.
Smaller clusters (1x1x1 and 2x1x1) have no or limited multicast opportunities and must issue redundant loads for shared operands.
These L2 bandwidth-bound configurations benefit from the bandwidth expansion, achieving up to a 2.64$\times$ speedup before reaching diminishing returns at roughly 8$\times$ L2 bandwidth. 

The second trend highlights a core trade-off: large clusters yield higher effective bandwidth, but imperfect fabrication severely impacts their occupancy. Even at 16$\times$ L2 bandwidth, a 4x2x1 configuration sustains only 90\% of 1x1x1 throughput. Because CGA mandates scheduling all CTAs of a cluster onto the same GPC, a GPC (18 SMs) fits at most two 8-SM clusters \cite{nvidia_h100}. Across eight GPCs, this requires 128 SMs, yet floorsweeping leaves the H100 with only 120 active SMs. Unlike small clusters that flexibly pack into surviving SMs, the rigid spatial constraint of large clusters leaves active cores idle, causing the 10\% throughput penalty.

The final trend applies to the FP8 kernel. Although its performance scales with
L2 bandwidth, it reaches only $\sim$1440~TFLOPS, or $\sim$73\% of peak dense FP8
throughput \cite{nvidia_h100}.
The shortfall does not come from the memory system: the kernel periodically promotes accumulator values to FP32, and this promotion executes on the SIMT pipeline, leaving the kernel bounded by SIMT throughput once L2 bandwidth is unconstrained.
\begin{tcolorbox}[
    colback=white!15,     
    colframe=black,      
    arc=3mm,             
    boxrule=0.5pt,       
    left=6pt, right=6pt, top=1pt, bottom=1pt, 
    width=\linewidth,    
]
{
\textbf{Takeaway \#1:}
\textit{L2 bandwidth binds only where inter-SM reuse is absent: single-CTA clusters gain up to 2.64$\times$ and saturate near 8$\times$ bandwidth, while 4x2x1 clusters recover the same traffic via TMA multicast and are compute-bound.
With bandwidth unconstrained, the residual gap to peak is architectural: cluster-to-GPC placement and floorsweeping cap occupancy at $\sim$90\%, and FP8 accumulator promotion shifts the limiter to SIMT throughput at $\sim$73\% of peak.}
}
\end{tcolorbox}


\subsection{Asynchronous Tax} \label{sync_overhead}

Hopper enforces strict execution ordering between producer (TMA) and consumer (WGMMA) warps via the \texttt{mbarrier.try\_wait} instruction to prevent shared memory data hazards. If the barrier condition is unmet, the warp executes a \texttt{nanosleep} instruction and retries later. This polling operates as a two-level mechanism: a hardware-level \texttt{try\_wait} that wakes immediately upon completion, which falls back to a secondary software-based polling loop upon timeout (described in Section \ref{comteporary_gpu}). The duration of this sleep interval dictates a critical trade-off: shorter intervals minimize wake-up latency but waste issue slots on frequent retries, whereas longer intervals conserve instruction bandwidth but risk stalling the warp after the barrier resolves. To quantify this asynchronous barrier overhead, we sweep the \texttt{nanosleep} cycle count, controlling how frequently the waiting warp wakes up to poll the \texttt{mbarrier}. We also include an \textit{oracle} point assuming a purely hardware-based system that can poll indefinitely, bounding the achievable performance while still enforcing the inter-warp \texttt{mbarrier} dependencies required for correctness.

\begin{figure}[h]
    \centering
    \includegraphics[width=\linewidth]{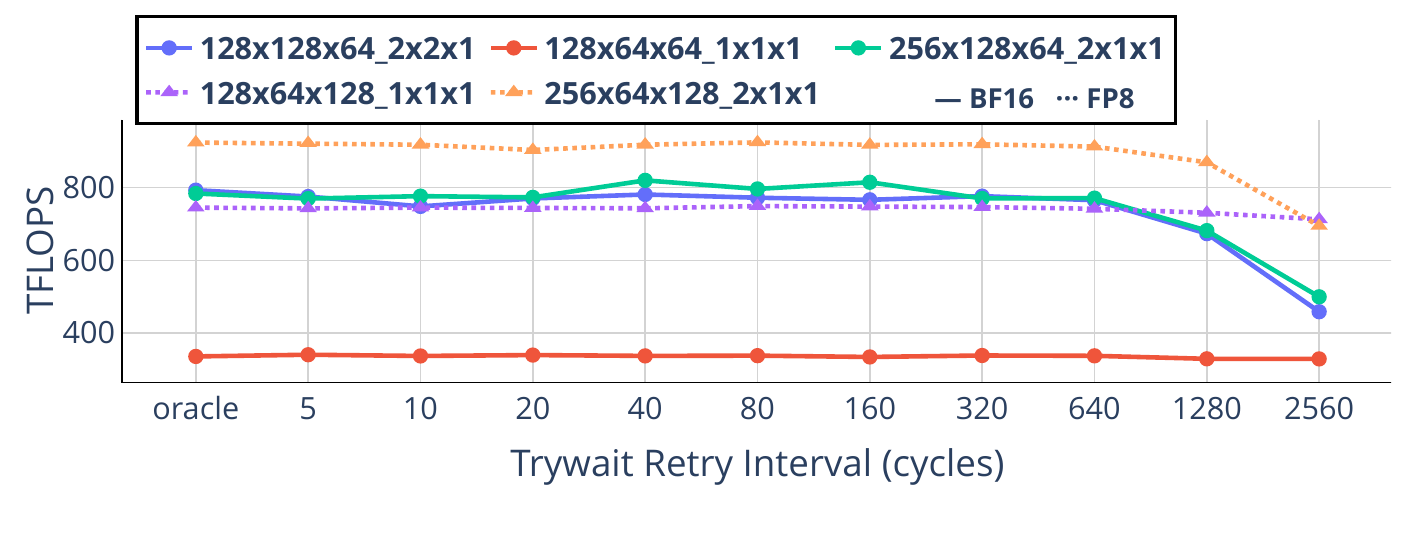}
    \caption{Simulated H100 TFLOPS for a 4096x4096x4096 dense GEMM in CUTLASS across a sweep of spinloop \texttt{NANOSLEEP} intervals. Kernel configurations are denoted as [Threadblock Tile Shape (MxNxK)] -- [Cluster Dimension].}
    \label{fig:async_retry_sweep}
\end{figure}

Figure~\ref{fig:async_retry_sweep} evaluates the same CUTLASS profiler kernel analyzed in Section~\ref{bw_expansion}. 
Across most configurations, the overhead of frequent barrier polling is negligible. Because heavy shared memory usage restricts occupancy to a single thread block (typically 12 warps) per SM, and because Hopper relies on long-latency asynchronous instructions (such as TMA loads and WGMMAs), the warp scheduler frequently experiences idle issue cycles with no eligible warps. 
Consequently, issuing additional \texttt{try\_wait} polling instructions simply consumes these idle cycles without degrading overall performance. 
In fact, execution time remains insensitive to polling frequency until the sleep interval becomes excessively large and causes warps to miss critical wake-up windows. Notably, the two 1x1x1 tile configurations exhibit zero degradation across the entire sweep. This occurs because 1x1x1 configurations do not utilize TMA multicast, an architectural requirement on Hopper to amplify effective L2 bandwidth and reach compute-bound peak TFLOPS. By remaining strictly memory-bound, the critical path of these kernels is occupied by asynchronous TMA memory transfers. Because the computation phases are short relative to the memory transfer times, the warps are not sensitive to precise scheduling; additional delays caused by inefficient polling or missed wake-up windows are effortlessly absorbed without extending the overall critical path.

\begin{tcolorbox}[
    colback=white!15,     
    colframe=black,      
    arc=3mm,             
    boxrule=0.5pt,       
    left=6pt, right=6pt, top=1pt, bottom=1pt, 
    width=\linewidth,    
]
\textbf{Takeaway \#2:} \textit{Barrier-polling overhead matters only when the warp scheduler is busy. In TMA-driven GEMM, long-latency asynchronous loads leave issue slots idle, and the extra \texttt{try\_wait} checks fill them at no cost to throughput until the sleep interval grows large enough to delay wake-up. 
}
\end{tcolorbox}

\section{Spatial Overhead}
\label{sec:micro-gpu}
\begin{figure}[h]
    \centering
    \includegraphics[width=\linewidth]{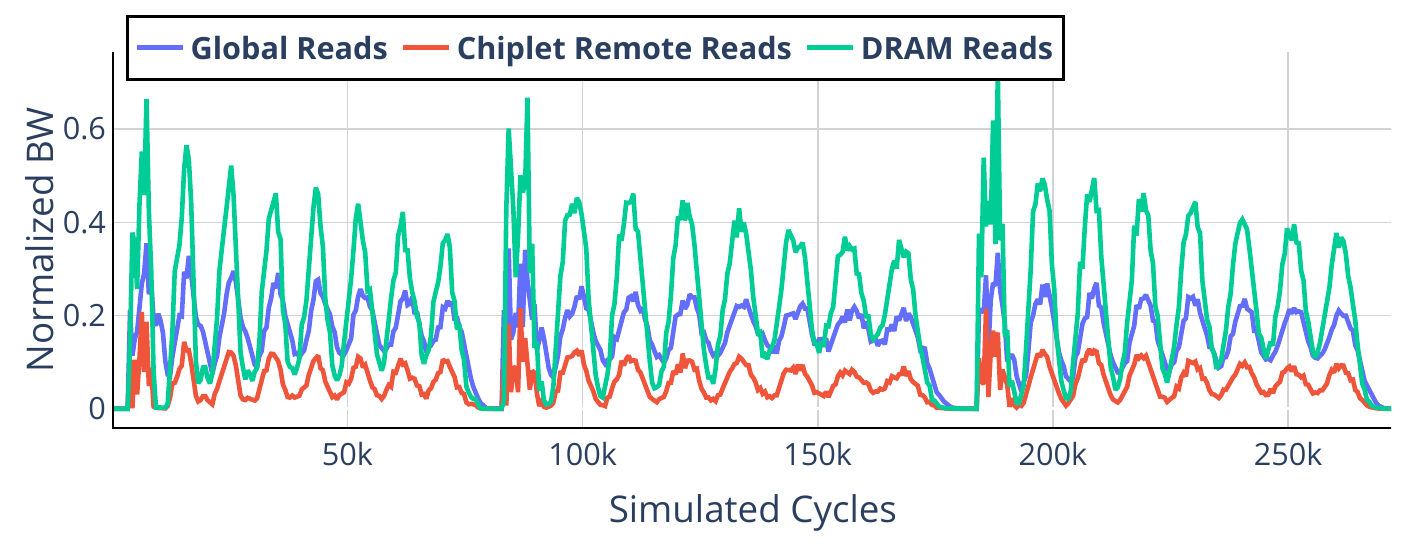}
    \caption{Simulated H100 (Section~\ref{evaluation}) bandwidth utilization for L2 global reads, DRAM reads, and chiplet remote reads for the MST graph workload \cite{lonestar-gpu}. Chiplet remote reads are normalized to the total L2 bandwidth (rather than remote load bandwidth) to illustrate their relative contribution to overall L2 traffic.}
    \label{fig:mst_l2_bw}
\end{figure}

To illustrate the potential effects of the chiplet architecture on the memory system, we first examine L2 bandwidth usage.
Figure \ref{fig:mst_l2_bw} shows the L2 read bandwidth for chiplet-local and inter-chiplet accesses in the Minimum Spanning Tree (MST) application, one of the benchmarks in the Lonestar benchmark suite \cite{lonestar-gpu}.
This benchmark computes a minimum spanning tree on a CSR graph.
We profiled three kernels within one iteration, which together perform per-iteration minimum cross-component edge discovery over the CSR graph, and which constitute the primary irregular read phase in the application.
Graph applications such as MST typically assign each CTA to a distinct subset of vertices or edges. However, CTAs still access the same global structures when traversing edges to neighboring nodes, leading to overlap in memory accesses across CTAs.
As shown in the figure \ref{fig:mst_l2_bw}, on average, 26\% of total accesses correspond to inter-chiplet traffic, indicating that CTAs mapped to SMs on different chiplets exhibit data reuse over shared global structures.

This highlights a critical microarchitectural trade-off in chiplet-based memory subsystems: because concurrent thread blocks frequently access shared structures, identical cache lines are redundantly replicated across per-chiplet L2 partitions, reducing effective global cache capacity. Table \ref{tab:mst_l2_hits} illustrates this penalty in MST, where the chiplet configuration increases L2 global Misses Per Kilo-Instruction (MPKI) by 86.46\% over a monolithic H100 architecture, driving a 24\% execution slowdown. We observe a similar penalty in ResNet-50 \cite{resnet}, where a cuDNN forward convolution executes 20.7\% faster on the monolithic L2 baseline due to a 20.1\% reduction in L2 misses. Although forward convolutions are typically compute-bound, their throughput relies on spatial data reuse; sacrificing effective capacity to replication induces unnecessary capacity misses. 

\begin{table}[h]
\caption{Simulation comparing a H100 partitioned L2 config versus a H100 GPU with monolithic L2 running MST\cite{lonestar-gpu}.}
\resizebox{\columnwidth}{!}{%
\begin{tabular}{lrrr}
\toprule
            Counter &    H100 & H100 Monolithic & Chipet/Monolithic \\
\midrule
     Kernel Latency & 272,134 &         219,456 &          +24.00\% \\
       L2 Global BW &  31.6\% &          38.2\% &          -17.25\% \\
L2 Global Load MPKI &  228.07 &          122.32 &          +86.46\% \\
            DRAM BW &  43.8\% &          29.7\% &          +47.37\% \\
\bottomrule
\end{tabular}
}
\label{tab:mst_l2_hits}
\end{table}


The LLM prefill phase proves resilient to these constraints. Despite processing large working sets that should theoretically induce cache thrashing, our evaluation demonstrates minimal impact.
To isolate this behavior, we evaluate a prefill pass using the FlashAttention-3 \cite{shah_flashattention3_fast_2024} kernel configured with a batch size of 10, an input sequence length of 544, and a context length of 12800 on Qwen-3 \cite{yang_qwen3_technical_2025}. This configuration yields a KV cache working set of approximately 50 MB. Although incorporating the query matrix slightly exceeds the H100 GPU's L2 capacity, this representative stress test shows that prefill cache utilization is largely immune to capacity fragmentation.

\begin{table}[h]
\caption{Simulation comparing a H100 partitioned L2 config versus a hypothetical H100 GPU with monolithic L2 running FA-3 prefill \cite{kwon_efficient_memory_2023,shah_flashattention3_fast_2024}. Global Reads only includes access to the local L2. Remote reads are normalized to total L2 bandwidth.}
\resizebox{\columnwidth}{!}{%
\begin{tabular}{lrrr}
\toprule
            Counter &      H100 & H100 Monolithic & Chiplet/Monolithic \\
\midrule
     Kernel Latency & 3,429,039 &       3,429,661 &            -0.02\% \\
       L2 Global BW &    45.5\% &          44.0\% &            +3.42\% \\
L2 Global Load MPKI &     59.54 &           49.45 &           +20.40\% \\
            DRAM BW &    12.7\% &          12.4\% &            +2.29\% \\
\bottomrule
\end{tabular}
}
\label{tab:prefill-chiplet}
\end{table}

Table \ref{tab:prefill-chiplet} details the simulated performance of this kernel modeled under an H100 configuration. Despite the reduced effective capacity of the partitioned design, overall kernel latency remains nearly identical. We observe a nominal difference in L2 global bandwidth. The small variance stems from the LRC. Because the LRC merge window is strictly bounded between initial request arrival and data return, timing shifts introduced by the partitioned design alter the coalescing rate, causing minor fluctuations in L2 accesses.

However, the partitioned design exhibits a 20.4\% increase in L2 misses. Because the prefill phase remains compute-bound, the additional memory accesses do not shift the performance bottleneck, resulting in no overall execution degradation. Nevertheless, this behavior carries energy implications. The additional misses require 20.4\% more data to traverse the inter-chip interconnect, increasing overall energy consumption \cite{khairy_localitycentric_data_2020}. HBM sector reads show only a marginal increase. This indicates that while local capacity is reduced, local misses frequently result in remote L2 hits. These remote hits satisfy the memory requests without generating additional HBM traffic. 
For the decode phase, because token generation is memory-bound and lacks inherent data reuse, it exhibits virtually no L2 locality. Consequently, the decode phase is insensitive to the effective capacity reductions introduced by the chiplet architecture.

\begin{tcolorbox}[
    colback=white!15,     
    colframe=black,      
    arc=3mm,             
    boxrule=0.5pt,       
    left=6pt, right=6pt, top=1pt, bottom=1pt, 
    width=\linewidth,    
]
{
\textbf{Takeaway \#3:} \textit{Aggressive data replication reduces the effective cache capacity available to locality-dependent applications, degrading performance for workloads such as graph analytics and CNNs. Because modern GPUs excel at hiding latency through memory-level parallelism, workloads are insulated from interconnect delays.}
}
\end{tcolorbox}

\section{Scale-Out Limits}
\label{sec:multi-gpu}
While multi-GPU execution is standard for scaling AI models, characterizing fine-grained execution bottlenecks remains challenging. For example, overlapping communication primitives with computation is critical for maximizing throughput, yet existing profilers only capture aggregate system-level performance. Furthermore, current multi-GPU simulators typically require custom benchmarks and cannot directly execute real-world PyTorch workloads.

To demonstrate how our framework overcomes these limitations, we present two case studies. 
First, we evaluate recommendation-system EmbeddingBag~\cite{pytorch_embbag} kernels in our multi-GPU simulator, identifying memory bandwidth saturation limits and proposing micro-architectural optimizations to push the performance envelope.
Second, we use GPUVision to profile NCCL and GEMM overlap; unlike traditional kernel-level profilers that report static summary metrics, our fine-grained time series exposes how Hopper's static tile assignment induces severe tail-latency bottlenecks during concurrent execution. 




\subsection{Embedding kernel in DLRMs} \label{dlrm}

The embedding lookup phase of DLRM models involves reading of several rows from embedding tables and performing a reduction operation on the values. The embedding bag operator is used to process a given table using the Pytorch CUDA backend kernel~\cite{pytorch_embbag}. While there are other variations of this kernel that perform different types of reduction operations (\textit{e.g.}, sum, mean, or max), the fundamental mechanism remains consistent across these flavors. Typically, native PyTorch assumes that all lookups corresponding to embedding tables are local to a GPU, meaning the data is stored in the HBM of the same GPU performing the computation. This assumption simplifies access patterns and ensures low-latency, high-throughput operations during embedding table lookups. The native PyTorch implementation of DLRM models is a memcpy orchestrated solution where the result of table lookups is transferred to the destination GPUs by cuda memcopies. It is possible to re-architect the embedding bag kernel to directly access data located on a neighboring GPU (via UVA) thus enabling a more programmer-friendly abstraction.

\begin{figure}[h]
    \vspace{-5pt}
    \centering
    \includegraphics[width=\linewidth]{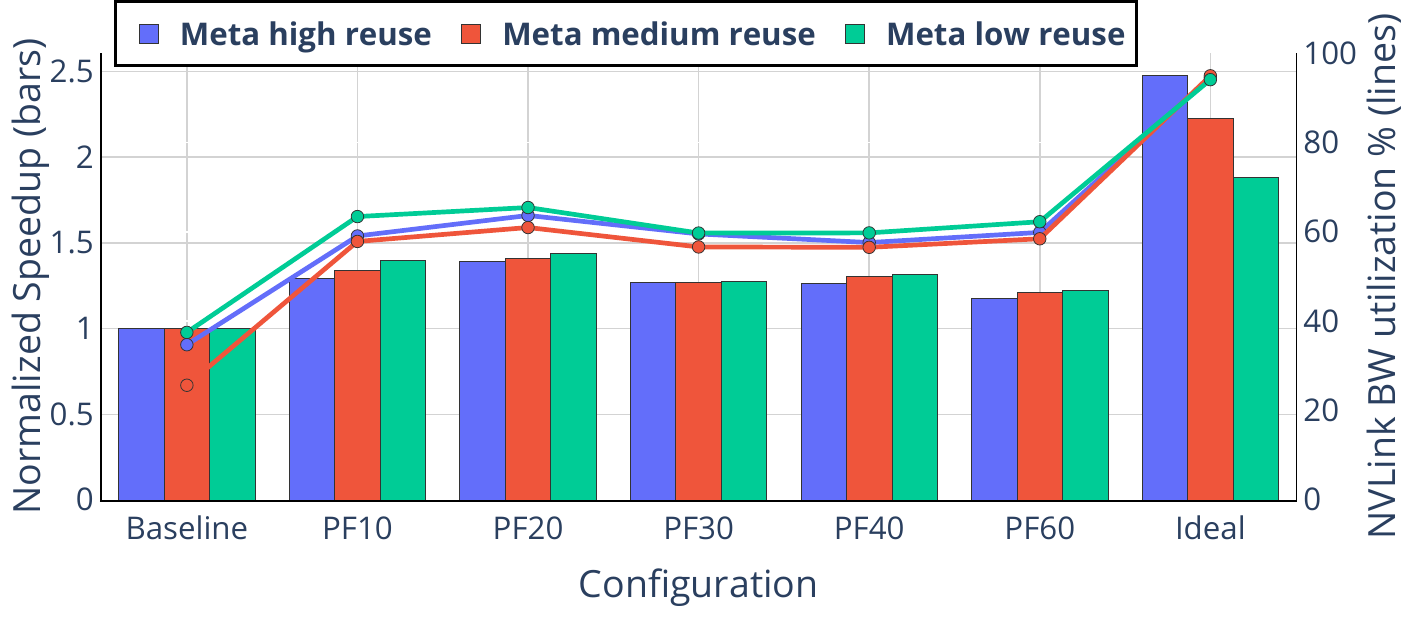}
    \caption{Simulated H100 (Section~\ref{evaluation}) multi-GPU performance and NVLink bandwidth results for the Baseline and Prefetch optimized EmbeddingBag kernels at various prefetch distances. The prefetch (PF) configurations are labeled as \texttt{PF\_<PF distance>}. Obtained using batch size of 1024 and embedding width of 80 on Meta production traces~\cite{metasyn}.}
    \vspace{-10pt}
    \label{fig:dlrm_pf}
\end{figure}

As the embedding kernel does a series of embedding table lookups, it is natural to ask if traditional compiler techniques such as prefetching and loop unrolling will help to improve the Memory Level Parallelism. Figure~\ref{fig:dlrm_pf} depicts the normalized speedup and the utilization of the inter-GPU interconnect bandwidth over the NVLink. Increasing the prefetch distance improves the speedup over the baseline due to better NVLink Utilization. However, increasing prefetch distances beyond 20 has limited benefits (and often a slight degradation in performance). This is because prefetching results in increased usage of on-chip resources such as registers and this limits the overall GPU occupancy (while the baseline runs 4 CTAs/SM, the occupancy drops down to 1-2 CTAs/SM at large prefetch distances). Thus, it is vital to ask if future GPU architectures can help prefetch across large distances without hurting occupancy and maximizing NVLink utilization as in the Ideal case.  

\begin{tcolorbox}[
    colback=white!15,     
    colframe=black,      
    arc=3mm,             
    boxrule=0.5pt,       
    left=6pt, right=6pt, top=1pt, bottom=1pt, 
    width=\linewidth,    
]
{
\textbf{Takeaway \#4:} \textit{The effectiveness of prefetching remote embedding lookups is constrained by CTA occupancy, which is itself bounded by scarce on-chip resources. This motivates architectures that can prefetch deeply without sacrificing occupancy}
}
\end{tcolorbox}


\subsection{NCCL/GEMM Overlapping}
Beyond simulation-driven exploration, our framework also incorporates GPUVision, a cycle-level profiling tool designed to characterize execution dynamics directly on physical hardware. To demonstrate the power of GPUVision on real silicon, we use it to evaluate the microarchitectural dynamics of NCCL/GEMM overlap, a common execution pattern in Fully Sharded Data Parallel (FSDP) configurations~\cite{zhao2023fsdp}. To guarantee deterministic scheduling and enable precise isolation of concurrency overheads, we construct a CUDA Graph~\cite{CUDA-Graph} that strictly partitions the workload into three phases: \Circled{1}~\textbf{Isolated NCCL}, a standalone AllGather; \Circled{2}~\textbf{Overlapped Execution}, the AllGather running concurrently with the GEMM; and \Circled{3}~\textbf{Isolated GEMM}, a standalone GEMM.

\begin{figure}[h]
    \vspace{-5pt}
    \centering
    \includegraphics[width=\linewidth]{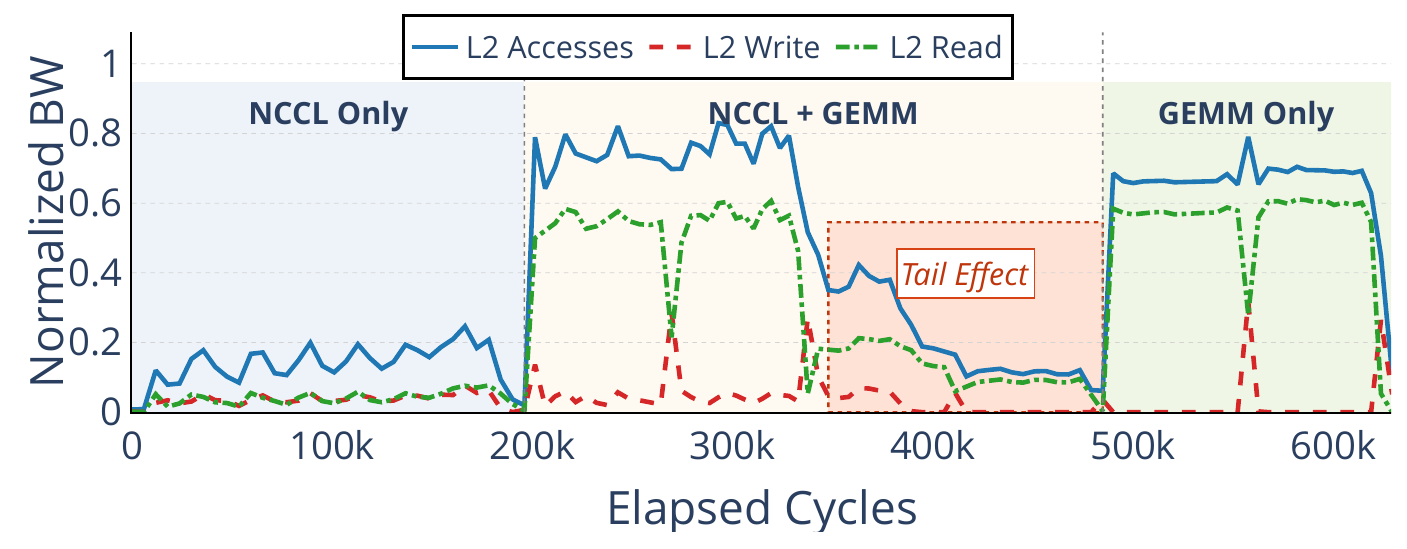}
    \caption{Silicon H200 GPUVision time series for L2 accesses, demonstrating the interference effects of cuBLAS GEMM ($2K \times 4K \times 4K$) and NCCL AllGather (8MB) kernels \cite{NCCL}.
    }
    \label{fig:nccl-gemm}
\end{figure}

As illustrated in Figure \ref{fig:nccl-gemm}, our high-resolution profiling uncovers a severe tail effect, exposing a limitation in Hopper’s static threadblock scheduling. To fully saturate NVLink bandwidth, the NCCL AllGather kernel persistently occupies a specific subset of the GPU's resources—typically 24 of the 132 available SMs. Concurrently, the overlapping computation is a highly optimized, cuBLAS GEMM, which is structurally designed to assume uniform access across the entire GPU. Because the kernel relies on a static tile assignment policy, the GEMM threadblocks preemptively mapped to the 24 NCCL-occupied SMs are forcibly deferred. Once the communication phase completes and yields those SMs, they must process their backlogged GEMM waves while the remaining 108 SMs sit entirely idle. This scheduling-induced tail effect demonstrates how occupancy-agnostic static assignment prevents dynamic work redistribution, leading to severe hardware underutilization during concurrent execution.
Exposing obscured bottlenecks like this tail effect is critical for optimizing distributed workloads. By providing the temporal resolution needed to visualize these rigidities, our profiling infrastructure enables architects to identify and mitigate hardware underutilization that standard kernel-level summary metrics would miss.

\section{{Evaluation}} \label{evaluation}

\begin{table*}[t]
\caption{H100 simulated kernel counts, application domains, and traced layer descriptions across ML workloads.}
\label{tab:workload-summary}
\resizebox{\textwidth}{!}{%
\begin{tabular}{lrll}
\hline
Suite                                                                         & Tot. Sim. Kernels & Domain                                                                          & Traced Layers/Kernels                                                                                                                                                                                                                                                                                                                               \\ \hline
BERT 256/512$^{\dagger}$ \cite{bert}                                          & 120               & \begin{tabular}[c]{@{}l@{}}NLP / Text\\ Classification\end{tabular}             & \begin{tabular}[c]{@{}l@{}}Encoder-only bidirectional attention, output projection, LayerNorm, dropout.\end{tabular}                                                                                                                                                                                                                              \\ \hline
ResNet-50 Batch Size 32/128 \cite{resnet}                                     & 550               & \begin{tabular}[c]{@{}l@{}}Vision / Image\\ Classification\end{tabular}         & \begin{tabular}[c]{@{}l@{}}Stages 1, 3, 4 (56$\times$56 to 7$\times$7); first and last Bottleneck blocks per stage.\end{tabular}                                                                                                                                                                                                                  \\ \hline
DLRMv2 \cite{dlrmv2}                                                          & 173               & \begin{tabular}[c]{@{}l@{}}RecSys / Click-\\ Through Rate\end{tabular}          & \begin{tabular}[c]{@{}l@{}}All four stages: \texttt{dense\_arch} (bottom FFN), \texttt{sparse\_arch} (embeddings),\\ \texttt{inter\_arch} (dot-product), \texttt{over\_arch} (top FFN).\end{tabular}                                                                                                                                              \\ \hline
3D U-Net \cite{unet3d}                                                        & 256               & \begin{tabular}[c]{@{}l@{}}Medical / 3D\\ Segmentation\end{tabular}             & \begin{tabular}[c]{@{}l@{}}Encoder entry, mid-level, bottleneck, two SkipConnection blocks, final decoder.\end{tabular}                                                                                                                                                                                                                           \\ \hline
RetinaNet \cite{retinanet}                                                    & 1{,}224           & \begin{tabular}[c]{@{}l@{}}Vision / Object\\ Detection\end{tabular}             & \begin{tabular}[c]{@{}l@{}}Early/late backbone, FPN, shared classification/regression heads.\end{tabular}                                                                                                                                                                                                                                         \\ \hline
Whisper Large V3 \cite{whisper}                                               & 968               & \begin{tabular}[c]{@{}l@{}}Audio / Speech\\ Recognition\end{tabular}            & \begin{tabular}[c]{@{}l@{}}Encoder layers 0, 31; decoder layers 0, 15, 31; vocabulary projection.\end{tabular}                                                                                                                                                                                                                                    \\ \hline
Stable Diffusion XL \cite{sdxl}                                               & 890               & Text-to-Image                                                                   & \begin{tabular}[c]{@{}l@{}}UNet down/mid/up paths, timestep FFN. One denoising step.\end{tabular}                                                                                                                                                                                                                                                 \\ \hline
FlashAttention \cite{shah_flashattention3_fast_2024, dao_flashattention2_faster_2023, zadouri_flashattention4_algorithm_2026}                        & 18                & Attention                                                                       & \begin{tabular}[c]{@{}l@{}}18 configs sweeping FlashAttention across Llama-2/3 head configurations\\ in prefill, decode, and mixed modes.\end{tabular}                                                                                                                                                                                          \\ \hline
CUTLASS Profiler \cite{thakkar_cutlass_2023}                                  & 98                & \begin{tabular}[c]{@{}l@{}}GEMM / Linear\\ Algebra\end{tabular}                 & \begin{tabular}[c]{@{}l@{}}4096x4096x4096 GEMM, BF16/FP8, different tiles.\end{tabular}                                                                                                                                                                                                                                                           \\ \hline
Llama 3.1 8B Training 256$^{\dagger}$ \cite{metallama_llama318binstruct_2024} & 240               & \begin{tabular}[c]{@{}l@{}}LLM / Language\\ Modeling\end{tabular}               & \multirow{5}{*}{\begin{tabular}[c]{@{}l@{}}Decoder-only with GQA and SwiGLU FFN. Early, middle, and late layers\\ sampled; attention and FFN traced independently. Llama traced under both\\ inference (vLLM) and training (includes backward-pass kernels). Mixtral uses\\ MoE FFN (8 experts, top-2) with FP8 quantization. Comprehensive single-block\\ kernel sampling (primary and auxiliary) validated to reproduce end-to-end\\ model latency within 3.06\% error.\end{tabular}} \\ \cline{1-3}
Llama 3.1 8B Inference 256/1024/2048$^{\dagger}$ \cite{metallama_llama318binstruct_2024} & 174   & \begin{tabular}[c]{@{}l@{}}LLM / Language\\ Modeling\end{tabular}               &                                                                                                                                                                                                                                                                                                                                                     \\ \cline{1-3}
DeepSeek-R1-Distill-Llama-8B 256/1024$^{\dagger}$ \cite{deepseekr1}           & 174               & LLM / Reasoning                                                                 &                                                                                                                                                                                                                                                                                                                                                     \\ \cline{1-3}
Qwen 2.5 7B 256/1024/2048$^{\dagger}$ \cite{qwen25}                           & 174               & \begin{tabular}[c]{@{}l@{}}LLM / Language\\ Modeling\end{tabular}               &                                                                                                                                                                                                                                                                                                                                                     \\ \cline{1-3}
Mixtral 256/1024$^{\dagger}$ \cite{mixtral}                                   & 384               & \begin{tabular}[c]{@{}l@{}}LLM / MoE\\ Inference\end{tabular}                   &                                                                                                                                                                                                                                                                                                                                                     \\ \hline
\multicolumn{4}{l}{\footnotesize $\dagger$ Numbers denote context length (input/output tokens).}                                                                                                                                                                                                                                                                                                                                                                                                                                        \\ \hline
\end{tabular}%
}
\end{table*}
While our primary evaluation targets the NVIDIA H100 (Hopper) as the contemporary industry standard, we demonstrate the robust, cross-architecture fidelity of Accel-Sim v2.0 by validating the framework against a comprehensive suite of high-impact workloads spanning three successive datacenter generations: Ampere, Hopper, and Blackwell\footnote{An NVBit v1.8 instrumentation bug on Blackwell rarely causes Uniform Register Zero (URZ) to return uninitialized non-zero values, preventing the simulation of certain workloads. Because this artifact is exceptionally rare, the affected kernels (roughly 20 out of 20,000 evaluated on B200) were dropped from our evaluation suite. The bug and the root cause have been reported to the NVBit developers.}.
\subsection{Workload}
We validate our simulator using a wide range of applications. Despite the surge of interest in LLMs, High-Performance Computing (HPC) workloads continue to drive significant architectural design. 
Because legacy HPC benchmarks cannot capture the execution dynamics of modern architectures, we augment them with a curated suite of NVIDIA-maintained libraries and high-performance physics engines, as detailed in Table \ref{tab:hpc-workload}. This extended evaluation includes modern primitives from the cuFFT \cite{cufft} and cuSolver \cite{cusolver2026} libraries to ensure broad coverage of contemporary hardware features.
Similarly, existing machine learning benchmark suites such as DeepBench and Polybench target legacy architectures and do not utilize new hardware features. To ensure our validation accurately reflects contemporary GPU execution, we abandon these outdated suites and instead compile a representative list of modern machine learning models.

\begin{table}[h]
\caption{HPC workload suites evaluated in this paper.}
\label{tab:hpc-workload}
\resizebox{\columnwidth}{!}{%
\begin{tabular}{lrl}
\hline
Suite           & Tot. Sim. Kernels & apps                                                                                                                            \\ \hline
Rodinia-3.1\cite{che_rodinia_2009}     & 5418              & \begin{tabular}[c]{@{}l@{}}b+tree, backprob, bfs, hotspot,\\ srad-v1, nn, needle, dwt2d, lud\end{tabular}   \\ \hline
Microbenchmarks\cite{khairy_accelsim_2020} & 16                & \begin{tabular}[c]{@{}l@{}}l1-lat, l1-bw, shd-lat, shd-bw, \\ l2-lat,l2-bw, mem-lat, mem-bw, \\ maxflops, multi-gpu-bw-lat\end{tabular} \\ \hline
CuFFT\cite{cufft}           & 24                & 3D Complex FFT, LTO filter                                                                                                      \\ \hline
CuGraph\cite{cugraph2026}         & 264               & Minimum spanning tree                                                                                                           \\ \hline
CuSolver\cite{cusolver2026}        & 183               & QR Factorization, LU Factorization                                                                                              \\ \hline
VPI\cite{VPI2026}             & 69                & \begin{tabular}[c]{@{}l@{}}Stereo disparity, Convolution2D, \\ Feature Detection\end{tabular}                                   \\ \hline
Newton-physics\cite{newton2026}  & 23206             & Ball simulation, Cartpole sim                                                                                                   \\ \hline
CUDA Samples\cite{cudasamples2026}    & 16                & DWT1D, FDTD3D           \\ \hline              
NCCL\cite{NCCL}    &  71              &  \begin{tabular}[c]{@{}l@{}}AllReduce, ReduceScatter,\\  AllGather, Broadcast\end{tabular}          \\ \hline 
\end{tabular}%
}
\end{table}
\begin{table}[]
\caption{Simulator validation MAPE vs. hardware (workloads: Tables~\ref{tab:workload-summary}, \ref{tab:hpc-workload}). N/A: unsupported.}
\label{tab:per-suite-mape}
\footnotesize 
\renewcommand{\arraystretch}{0.95} 
\begin{tabular*}{\columnwidth}{@{\extracolsep{\fill}}|l|r|r|r|}
\hline
Suite                   & \multicolumn{1}{l|}{A100} & \multicolumn{1}{l|}{\textbf{H100}} & \multicolumn{1}{l|}{B200} \\ \hline
Rodinia-3               & 19.3                      & \textbf{16.1}                      & 7.6                       \\ \hline
GPU Microbenchmark      & 15.9                      & \textbf{11.1}                      & 16.7                      \\ \hline
cuFFT                   & 26.5                      & \textbf{17.2}                      & 2.5                       \\ \hline
cuSOLVER                & 23.0                      & \textbf{19.7}                      & 37.5                      \\ \hline
cuGraph                 & 19.0                      & \textbf{7.5}                       & 0                         \\ \hline
Newton Physics          & 10.7                      & \textbf{9.0}                       & 3                         \\ \hline
VPI                     & 9.3                       & \textbf{16.3}                      & 7.8                       \\ \hline
CUDA Samples            & 32.5                      & \textbf{10.5}                      & 5.7                       \\ \hline
BERT                    & 13.2                      & \textbf{3.7}                       & 14                        \\ \hline
ResNet-50               & 8.1                       & \textbf{14.4}                      & 8                         \\ \hline
DLRMv2                  & 3.3                       & \textbf{15.4}                      & 12                        \\ \hline
3D U-Net                & 11.9                      & \textbf{13.2}                      & 10                        \\ \hline
RetinaNet               & 2.1                       & \textbf{5.4}                       & 2                         \\ \hline
Whisper Large V3        & 11.6                      & \textbf{2.3}                       & 3                         \\ \hline
Stable Diffusion XL     & 13                        & \textbf{9.4}                       & 3                         \\ \hline
FlashAttention        & 15.2 (FA-2)               & \textbf{15.4 (FA-3)}               & 11.7 (FA-4)               \\ \hline
CUTLASS (sm\_90) & N/A                       & \textbf{21.2}                      & N/A                       \\ \hline
Llama 3.1 8B Training   & 24                        & \textbf{16.1}                      & 20                        \\ \hline
Llama 3.1 8B Inference  & 17.6                      & \textbf{19.6}                      & 14                        \\ \hline
DeepSeek R1 8B          & 18.1                      & \textbf{8.0}                       & 5                         \\ \hline
Qwen 2.5 7B 256/2k      & 6.9                       & \textbf{21.1}                      & 3.67                      \\ \hline
Mixtral                 & 18.03                     & \textbf{21.6}                      & 2.5                       \\ \hline
NCCL                    & 18.7                      & \textbf{17}                        & N/A                       \\ \hline
\textbf{All}                     & \textbf{15.4}                      & \textbf{13.5}                      & \textbf{8.9}                       \\ \hline
\end{tabular*}
\end{table}

\Cref{tab:workload-summary} summarizes the machine learning workload suites used in our evaluation. Our profiling methodology is highly extensible and straightforward to configure; by utilizing standard vLLM and Hugging Face Transformers APIs, any model hosted on Hugging Face that supports local execution can be set up and traced with just a few lines of script. The Traced Layers column in the table details the specific layers and sub-layers selected for each model. For LLM inference workloads (Llama-3.1-8B \cite{metallama_llama318binstruct_2024}, Qwen2.5-7B \cite{qwen25}, Mixtral~8x7B \cite{mixtral}, DeepSeek-R1-Distill-Llama-8B \cite{deepseekr1}), we trace the prefill pass followed by one decode step at multiple context lengths. Mixtral additionally uses FP8 quantization to fit on a single GPU. 
For Llama-3.1-8B training, we trace one full training step (forward pass, backward pass, and optimizer step) with a sequence length of 256 using Hugging Face Transformers. BERT-base \cite{bert} processes a single forward pass (encoder-only, no autoregressive decoding) with batch size 32 at sequence lengths 256 and 512. ResNet-50 \cite{resnet} uses batch sizes of 32 and 128 with $224{\times}224$ ImageNet inputs. DLRMv2 \cite{dlrmv2} uses batch size 256 with 26 embedding tables. 3D-UNet \cite{unet3d} processes a single $128^3$ volume. RetinaNet \cite{retinanet} uses batch size 4 with $800{\times}800$ images. Whisper Large V3 \cite{whisper} processes 30 seconds of audio. SDXL \cite{sdxl} generates a single $1024{\times}1024$ image with one denoising step, as the UNet architecture is identical across denoising steps. 
Kernel counts shown in \Cref{tab:workload-summary} and \Cref{tab:hpc-workload} are based on the H100.

\subsection{Correlation} \label{validation}

\begin{figure*}[h]
    \centering
    \includegraphics[width=\linewidth]{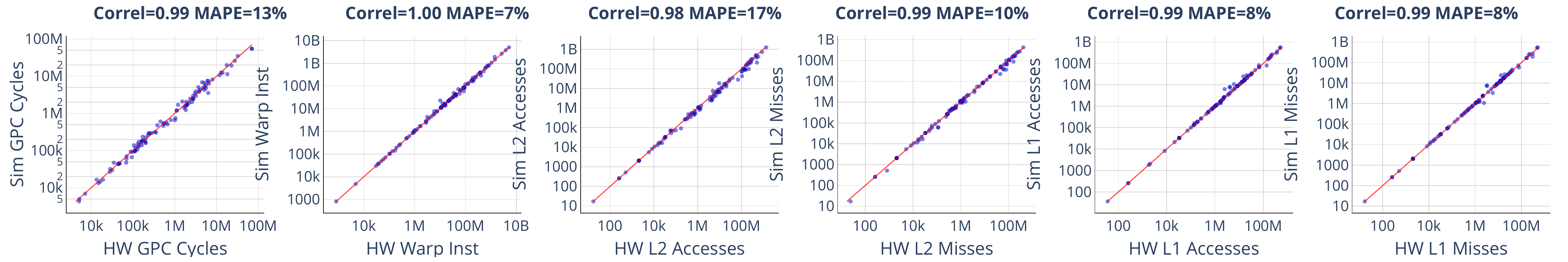}
    \caption{Correlation of key performance metrics between the Accel-Sim 2.0 and the NVIDIA H100 GPU. The results demonstrate high fidelity for the diverse workload suites described in Table \ref{tab:workload-summary} and Table \ref{tab:hpc-workload}, totaling more than 34,000 kernel instances.}
    \label{fig:correlation}
\end{figure*}

\begin{figure}[h]
    \centering
    \includegraphics[width=\linewidth]{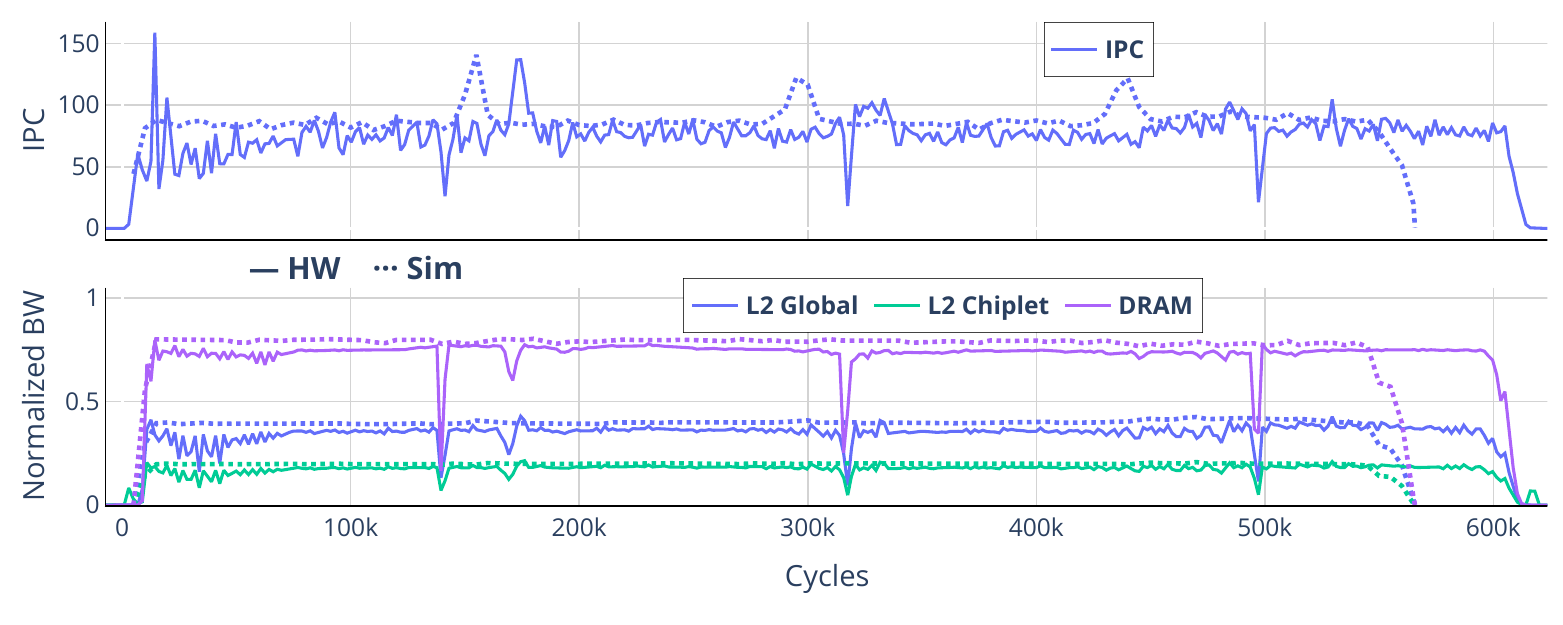}
    \caption{H100 Cycle-level correlation enabled with GPUVision. The workload is FlashAttention-3 running decode.}
    \label{fig:fa3_timeseries}
    \vspace{-10pt}
\end{figure}

Figure \ref{fig:correlation} illustrates the Pearson correlation and mean absolute percent error (MAPE) of Accel-Sim 2.0 running H100 config as shown in Table \ref{tab:sim_config} across the workloads detailed in Table \ref{tab:workload-summary} and Table \ref{tab:hpc-workload}. Overall, the simulator achieves a 0.99 correlation and a 13.5\% MAPE compared to NVIDIA H100 hardware. Table \ref{tab:per-suite-mape} shows per-suite MAPE of the 3 datacenter cards evaluated.
Among all evaluated metrics, L2 cache traffic exhibits the highest MAPE. This discrepancy is driven by the LRC. Because the LRC relies on a strict temporal merge window, minor, non-deterministic timing variations in real hardware cause threadblocks to miss coalescing opportunities. Consequently, physical L2 traffic inherently fluctuates across identical kernel runs.
During validation, we observed that certain hardware performance counters on newer GPU generations can be unreliable. While the \textit{ltcfabric} L2 cache metrics in NVIDIA Nsight Compute are intended to measure inter-chiplet L2 accesses, we found these values to be unstable. 
To correlate L2 chiplet accesses, we use GPUVision to profile \path{lts__t_sectors_srcunit_tex.sum}, which isolates the L2 sectors originating from the L1 data cache, and subtract this from the total L2 sectors accesses (\path{lts__t_sectors.sum}) to derive the chiplet accesses.
Figure~\ref{fig:fa3_timeseries} shows the resulting cycle-level correlation, confirming that the simulator tracks the memory subsystem's behavior accurately.
This high-fidelity approach allows researchers to precisely pinpoint system bottlenecks and accelerates the architectural research process. The simulator accurately reflects hardware dynamics, providing a high degree of confidence in the simulated results.

\section{Related Work}
\textbf{Cycle-Level GPU Simulation.} Cycle-level GPU simulation has evolved from functional PTX modeling in GPGPU-Sim \cite{bakhoda_analyzing_2009} to SASS-level trace-driven modeling in Accel-Sim \cite{khairy_accelsim_2020}, which forms our foundational building block. While simulators like MGPUSim \cite{sun_mgpusim_enabling_2019} and gem5-APU \cite{power_gem5gpu_2015} target AMD GCN3, NVIDIA's proprietary NVAS framework \cite{villa_need_2021} models high-fidelity silicon; we adopt NVAS's spinloop-detection methodology into our open-source tracer to ensure deterministic replay. Huerta et al. \cite{huerta_dissecting_2025} recently reverse-engineered consumer-grade core pipelines with 99\% correlation. We diverge by explicitly targeting datacenter-class architectures (modeling B200 and H100 features like TMA multicast and 2-CTA execution from Section \ref{bw_expansion}) and the asynchronous AI programming paradigms (TMA, WGMMA, and \texttt{mbarriers}) that drive them.

\textbf{Microbenchmarking and NUMA Architectures.} Extensive microbenchmarking literature has dissected the Hopper \cite{luo_dissecting_nvidia_2025} and Blackwell \cite{jarmusch_microbenchmarking_nvidias_2026} memory hierarchies and execution pipelines. Beyond single-core execution, Jin et al. \cite{jin_uncovering_real_2024} uncovered severe NUMA effects in single-package GPUs via NoC analysis, which Choudhary et al. \cite{choudhary_optimizing_attention_2025} mitigated through software-level memory mapping for attention mechanisms. While LADM \cite{khairy_localitycentric_data_2020} evaluated NUMA effects in older multi-GPU HPC systems, our work natively models disaggregated NUMA topologies within a cycle-level framework to study their direct impact on phase-specific LLM execution.

\textbf{Scale-Out Simulation and Analytical Modeling.} Scale-out frameworks like ASTRA-Sim \cite{rashidi_astrasim_enabling_2020}, Astra-Sim 3.0 \cite{astrasim3} (incorporating MSCCL++ \cite{msccl++}), and TrioSim \cite{li_triosim_lightweight_2025} evaluate distributed DNN workloads by abstracting away SM microarchitecture, whereas we model both inter- and intra-core behaviors. Similarly, numerous analytical models and high-level simulators target LLM serving dynamics \cite{cho_llmservingsim_hw_2024, cho_llmservingsim20_unified_2025, parashar_timeloop_2019, agrawal_vidur_largescale_2024, pope_efficiently_scaling_, yuan_llm_inference_2024}. An extensive body of analytical and ML-assisted GPU performance models \cite{hong_analytical_2009, hong_integrated_2010, baghsorkhi_adaptive_2010, zhang_quantitative_2011, sim_gpuperf_2012, huang_gpumech_2014, wu_gpgpu_2015, wang_mdm_2020, lee_gcom_2022, seyyedaghaei_scalemodel_2024, cha_gcstack_2025, zhang_pipeweave_2026}, alongside recent end-to-end LLM inference predictors \cite{cao_amali_2025, li_pathforward_2023, zhang_llmcompass_2024, lee_neusight_2025}, enable rapid design space exploration. However, these macro-level tools operate at an abstraction level that precludes cycle-level analysis of microarchitectural contention, asynchronous barrier coordination, and hardware-orchestrated data movement.

\section{Conclusion}
To overcome the simulation limitations constraining contemporary GPU evaluation, this paper introduces Accel-Sim v2.0, a cycle-level framework natively modeling Hopper and Blackwell generation architectures. Validated against both Hopper and Blackwell silicon, our framework consistently achieves a 99\% Pearson correlation across both architectures, alongside mean absolute errors of 13.5\% and 8.9\%, respectively. By restoring cycle-level visibility, we expose critical microarchitectural bottlenecks in contemporary AI workloads: specifically, we demonstrate that while hardware-orchestrated data movement masks synchronization latency, performance remains severely constrained by intra-GPU NUMA cache replication penalties and register-bound inter-GPU prefetching. By uncovering these trade-offs, Accel-Sim v2.0 provides the community with the essential infrastructure needed to navigate the physical constraints of the LLM era.



\bibliographystyle{IEEEtran}
\bibliography{references}

\end{document}